# The Hermite and Fourier transforms in sparse reconstruction of sinusoidal signals

Maja Vešović, Valentina Konatar
University of Montenegro
Faculty of Electrical Engineering
DzordzaVasingtona bb, 20000 Podgorica, Montenegro

*Abstract*— **The paper observes the Hermite and the Fourier Transform domains in terms of Frequency Hopping Spread Spectrum signals sparsification. Sparse signals can be recovered from a reduced set of samples by using the Compressive Sensing approach. The under-sampling and the reconstruction of those signals are also analyzed in this paper. The number of measurements (available signal samples) is varied and reconstruction performance is tested in all considered cases and for both observed domains. The signal recovery is done using an adaptive gradient based algorithm. The theory is verified with the experimental results.**

## I. Introduction

The focus of this paper is on two commonly used transform domains in signal processing, and their possibility to provide sparse representation for specific types of signals. Sinusoidal signals are used in the experiments. Specifically, we observed the Frequency Hopping Spread Spectrum (FHSS) signals [1]-[6]. Sparse representation of those signals is of interest in Compressive Sensing (CS) approach [7]-[21]. Namely, having a sparse representation in a certain transform domain, the signal can be represented using fewer samples compared to the traditional approach based on the Sampling theorem. Therefore, in our paper we provided an analysis to decide which domain is better for FHSS signal representation in terms of sparsity.

FHSS signals consist of sinusoidal components of short duration and founds its usage in wireless communications. The FHSS modulation technique uses pseudorandom sequence to determine the frequency at which carrier will appear, causing spreading the spectrum of an unmodulated signal.

In recent years, the CS has its growth in many fields of signal processing and in a number of real applications. It is also used in the field of communications. The CS approach assures signal reconstruction using a small set of available signal samples, and provides satisfactory reconstruction precision if certain conditions are satisfied. First condition is sparsity of the observed signal in certain domain. Second condition is related to the acquisition procedure and it is satisfied if the available samples are randomly distributed. Recovering of the signal using a small set of available coefficients is a demanding task. It is based on the powerful mathematical apparatus and the optimization algorithms.

Let us now discuss the choice of transform domains to be tested for sparsification. Since the Hermite functions allow us obtaining good localization of the signals in both, the signal and transform domains, the Hermite transform (HT) is widely used in image and signal processing applications [9], [22]. HT functions are similar in shape to the FHSS components so they are chosen as a starting basis to be tested. In this paper, we also worked with Discrete Fourier Transform (DFT) domain. DFT is one of the commonly used transform domains in signal processing and analysis. Having in mind that we observe sinusoidal signals, and that sinusoid is represented with a peak in the DFT domain, this seemed to be a good choice for testing.

We compared reconstruction of signals by using both, HT and DFT domains. The number of measurements (available signal samples) is varied and reconstruction performance is tested in all considered cases and for both observed domains.

The paper is organized as follows: Section II is the theoretical background on CS technique. In Section III, Hermite and Fourier domain are described. The experimental results are discussed in Section IV. Concluding remarks are given in Section V.

## II. Theoretical Background

In order to be applied to the certain signals and in specific applications, some conditions related to the CS technique has to be satisfied. As it is already mentioned, signal should have a sparse representation in certain transform domain. Let us mathematically describe those conditions. If the length of the observed signal is $N$, then only $S$ transform coefficients should have non-zero values, where $S<N$. For such a signal we say that it is compressible and the information about the signal is concentrated into the $S$ largest coefficients. Another requirement that should be satisfied is an incoherence. Signals that have sparse representation in the transformation domain must have a considerably denser representation in the domain in which are represented. The incoherence condition is satisfied if random selection of the signal samples is performed.

If they are represented in the proper basis, most of the real signals can be considered as sparse. The signal **x** with $N$ samples can be represented as a linear combination of the basis vectors:

$$\mathbf{x} = \sum_{i=1}^{N} X_i Y_i, \quad (1)$$

where **Y** is transformation matrix and **X** is signal in **Y** domain.

We can say that the signal has sparsity level $S$, if only $S$ coefficients have nonzero values. Measurements $M$ are samples of signal taken from the domain in which the signal has dense representation. The number of measurements satisfies the relation $M<N$. The measurement matrix **θ** models random selection of the signal samples. It has to be incoherent with transformation matrix **Y**, in order to assure the successful signal reconstruction. Measurement matrix **θ** should be chosen in a way that provides as small as possible the coherence between matrices **θ** and **Y**. Small coherence means that it is possible to take small number of signal measurements and reconstruct the signal with great accuracy. The measurement vector can be defined by using the following relation:

$$\mathbf{y} = \mathbf{\theta x} = \mathbf{\theta X Y} = \mathbf{AY}, \quad (2)$$

where **A** is the measurement matrix ($M \times N$).

CS uses some mathematical algorithms for error minimization. In this paper we used gradient based algorithm [13]. It performs a direct search over all missing samples of the signal. If the values of every missing sample are located in the range of $-M$ to $M$ then for every missing sample the algorithm performs a search over all possible values in the given range by taking any given step. Larger steps are taken in the first few ruff approximations. As we get close to the true value of the missing samples, the step size is lowered in order to achieve a desired precision. This algorithm uses adaptive variable step in order to perform the reconstruction in a small number of iterations.

III. THE HT AND DFT IN TERMS OF SIGNAL SPARSITY

Frequency Hopped Spread Spectrum signals (FHSS) are communication signals with very close components. They are sinusoidally modulated signals that belong to the spread spectrum modulations. The spread spectrum (SS) modulation is a commonly used modulation technique in communications. It has many desirable properties such as robustness to inter-symbol interference, noise, jamming and other environmental factors. This modulation technique spreads the frequency spectrum of data-signal, producing the signal with much higher bandwidth than before. There are two common types of spread spectrum modulations: direct sequence spread spectrum (DSSS) and frequency hopping spread spectrum. Our focus in this paper are on the FHSS modulated signals.

The HT is observed for representation of the FHSS signals having in mind similarity in shape between Hermite expansion functions and the FHSS signal components. The definition of Hermite functions is given by:

$$\Psi_p(n) = \left(e^{-n^2} Hp(n)\right) / \left(\sqrt{2^p \, p! \sqrt{\pi}}\right) \quad (3)$$

where $Hp(n)$ is the $p$-th order Hermite polynomial. Also, the definition can be done using the following recursion formula:

$$\Psi_0(n) = \frac{1}{\sqrt[4]{\pi}} e^{-n^2/2}, \quad \Psi_1(n) = \frac{\sqrt{2}n}{\sqrt[4]{\pi}} e^{-n^2/2}$$

$$\Psi_p(n) = n\sqrt{\frac{2}{p}} \Psi_{p-1}(n) - \sqrt{\frac{p-1}{p}} \Psi_{p-2}(n), \forall_p \geq 2 \quad (4)$$

A continuous-time signal $f(i)$ is defined by using $N$ Hermite functions as:

$$f(i) = \sum_{p=0}^{N-1} C_p \Psi_p(i), \quad (5)$$

where $C_p$ denotes Hermite expansion coefficients. To calculate the Hermite expansion coefficients the Gauss-Hermite quadrature technique can be used, defined as:

$$C_p \approx \frac{1}{M} \sum_{m=1}^{M} \frac{\Psi_p(n_m)}{\left[\Psi_{M-1}(n_m)\right]^2} f(n_m). \quad (6)$$

Considering this technique the direct Hermite transform can be defined in a matrix form as follows:

$$\mathbf{C} = \mathbf{Hf}, \quad (7)$$

where **C** and **H** are vector of Hermite coefficients and Hermite transform matrix respectively. The inverse transform reads:

$$\mathbf{f} = \mathbf{\Psi C}, \quad (8)$$

where **Ψ** stands for inverse Hermite matrix $\mathbf{\Psi} = \mathbf{H}^{-1}$.

Having in mind that sinusoidal signals, in Fourier transform domain, are represented by a peak at the signal frequency, the second choice in our analysis is the DFT. If the matrix **Ψ** stands for the inverse Fourier transform matrix, then it is defined as follows:

$$\mathbf{\Psi} = \frac{1}{N}\begin{bmatrix} 1 & 1 & \cdots & 1 \\ 1 & e^{j\frac{2\pi}{N}} & \cdots & e^{(N-1)j\frac{2\pi}{N}} \\ 1 & e^{2j\frac{2\pi}{N}} & \cdots & e^{(N-1)2j\frac{2\pi}{N}} \\ \cdots & \cdots & \cdots & \cdots \\ 1 & e^{(N-2)j\frac{2\pi}{N}} & \cdots & e^{(N-1)(N-2)j\frac{2\pi}{N}} \\ 1 & e^{(N-1)j\frac{2\pi}{N}} & \cdots & e^{(N-1)(N-1)j\frac{2\pi}{N}} \end{bmatrix} \quad (9)$$

IV. EXPERIMENTAL RESULTS

We considered following FHSS signal:

$$x = e^{-j20\pi t(1:N/3)} + e^{j14\pi t(N/3:2N/3)} + e^{-j4\pi t(2N/3:N)}, \quad (10)$$

where $t=1:1/100:1-1/100$, and $N$ is the signal length. The total signal length is 600 samples and signal has 3 components - hops. Duration of every hop is the same but they have different frequencies. The considered three-component signal in time domain, and its DFT and HT domains are shown in Fig. 1. It can be seen that signal has better sparsity in DFT domain. Reconstruction is tested using different number of samples and both domains – HT and DFT. The results for DFT case are shown in Fig.2. In this particular case the reconstruction is done from 180 samples (30% the total signal length) and 480 (80% the total signal length), respectively.

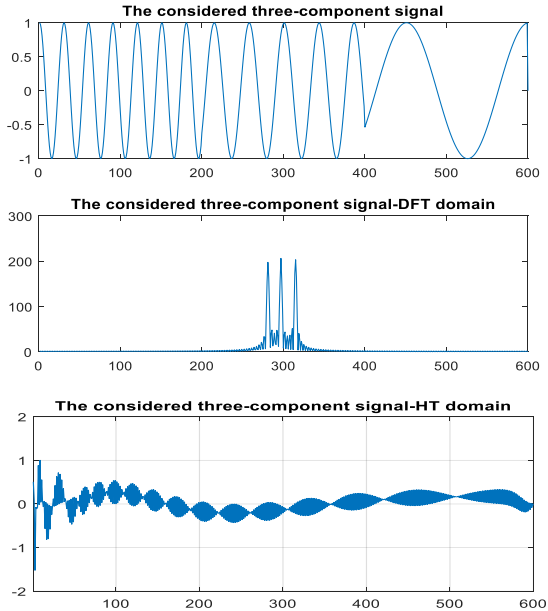

Figure 1. The considered three-component signal: first row - time domain, second row – DFT, third row– HT domain

The reconstruction results by using the HT domain are shown in Fig. 3, for the same number of available samples as in the DFT case. The signal has worse sparsity in the HT domain, what can be seen from the Fig.1. Therefore, the reconstruction results are worse compared to the case where the DFT is used as a sparse basis.

Increasing the number of measurements in both DFT and HT domain improves the sparsity and produces smaller error between the original and the reconstructed signal. Still, the HT domain shows worse reconstruction results compared to the DFT, even in the case when 80% of the samples are available. Fig. 4 shows original and reconstructed signals in time domain, using 30% and 80% of the available samples and DFT as a sparse basis. It can be seen that signal reconstructed from 80% of the signal samples produces almost negligible error.

However we cannot say that the reconstruction from the same number of measurements in HT domain gives the same results. On the contrary, HT fails to give reliable reconstruction, even in the case when large number of measurements (80%) is used. The reconstruction results are using HT domain are shown in Fig. 5. The reconstructed signal is almost uncrecognizable when using 30% of measurements and HT as a sparse basis.

The previous results are also confirmed in Fig.6, by calculating MSE for both DFT (red line) and HT (blue line) for different number of measurements. It is demonstrated that HT for 80% measurements produces almost the same MSE as DFT for 30% of measurements.

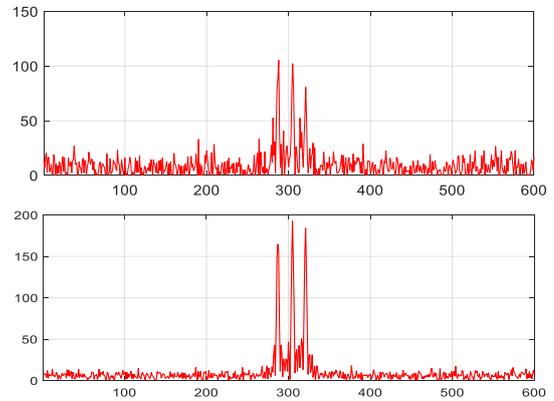

Figure 2. Reconstructed signal in DFT domain using: 180 samples or 30% of the signal length –first row; 480 samples or 80% of signal length-second row

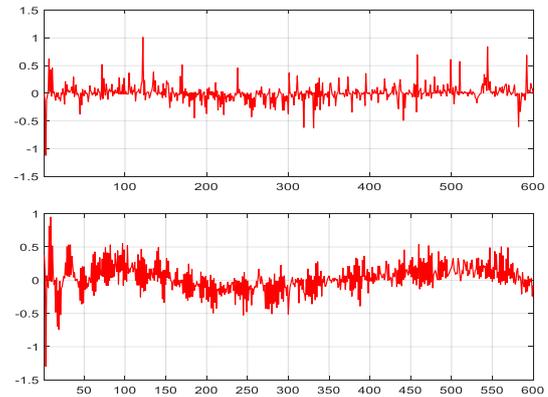

Figure 3. Original signal – HT domain,180 samples or 30% of signal length(first row);Original signal – HT domain – 480 samples or 80% of signal length(second row)

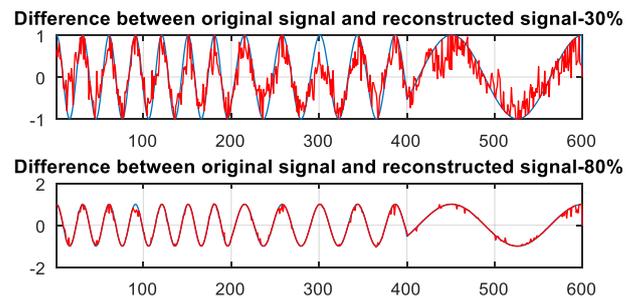

Figure 4. Time domain of the original and reconstructed signals by using 30% and 80% of the signal samples for the reconstruction and DFT as a sparse basis

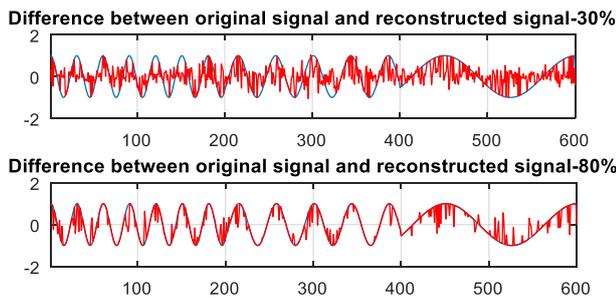

Figure 5. Time domain of the original and reconstructed signals by using 30% and 80% of the signal samples for the reconstruction and HT as a sparse basis

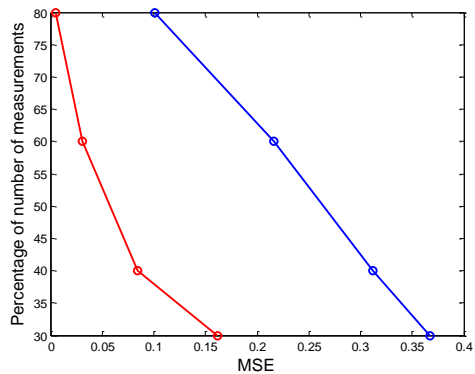

Figure 6. MSE for different number of measurements for DFT (red line) and HT (blue line)

V. CONCLUSION

The comparison between HT and DFT domains used for spread spectrum modulated signals sparsification is observed in the paper. The CS reconstruction using different number of signal measurements and both considered transformations as a sparse basis, is tested. The gradient based algorithm is used for the reconstruction. The results showed that the DFT domain is a better choice for the sparse representation of the multicomponent spread spectrum modulated signals. The HT may be a better choice if separated signal components are observed, and it can be a topic for future research. The paper demonstrates that the HT fails to give a reliable reconstruction, even when a large number of measurements is used. The results are verified by measuring the MSE between original and reconstructed signal for HT and DFT domains and for different number of measurements.

REFERENCES

[1] S. Barbarossa, J. L. Krolok, "Adaptive time-varying cancellation of wideband interferences in spread-spectrum communications based on time-frequencydistributions," IEEE Transactions on Signal Processing, Vol. 47, No. 4, Apr. 1999, pp. 957-965.

[2] M. Brajovic, A. Draganic, I. Orovic, S. Stankovic, "Sparse Representation of FHSS Signals in the Hermite Transform Domain," Telfor Journal, accepted for publication, 2017.

[3] M. Ghavami, L. B. Michael, R. Kohno, "Ultra Wideband Signals and Systems in Communication Engineering", Second Edition, John Wiley & Sons, Ltd, ISBN 978-0-470-02763-9 (HB), February 2007.

[4] A. Draganic, I. Orovic, S. Stankovic, X. Li, Z. Wang, "An approach to classification and under-sampling of the interfering wireless signals," Microprocessors and Microsystems, vol. 51, pp. 106-113, June 2017.

[5] M. Gandetto, M. Guainazzo, C. S. Regazzoni, "Use of time-frequency analysis and neural networks for mode identification in a wireless software-defined radio approach", EURASIP Journal on Applied Signal Processing, Vol. 2004, pp. 1778-1790, 2004.

[6] A. Draganic, I. Orovic, S. Stankovic, "Spread-spectrum-modulated signal denoising based on median ambiguity function," 59th International Symposium ELMAR-2017, Zadar, Croatia, 2017.

[7] D. Donoho: "Compressed sensing," IEEE Trans. on Information Theory, 2006, 52, (4), pp. 1289 – 1306.

[8] I. Volaric, V. Sucic, S. Stankovic, "A Data Driven Compressive Sensing Approach for Time-Frequency Signal Enhancement," Signal Processing, Volume 141, December 2017, Pages 229-239.

[9] S. Stankovic, I. Orovic, E. Sejdic, "Multimedia Signals and Systems: Basic and Advance Algorithms for Signal Processing," Springer-Verlag, New York, 2015.

[10] E. Candes, J. Romberg, "l1-magic: Recovery of Sparse Signals via Convex Programming", October 2005.

[11] E. Sejdic, I. Orovic, S. Stankovic, "Compressive sensing meets time-frequency: An overview of recent advances in time-frequency processing of sparse signals," Digital Signal Processing, accepted for publication, 2017.

[12] S. Stankovic, I. Orovic, LJ. Stankovic, A. Draganic, "Single-Iteration Algorithm for Compressive Sensing Reconstruction," Telfor Journal, Vol. 6, No. 1, pp. 36-41, 2014.

[13] G. Pope, "Compressive Sensing: a Summary of Reconstruction Algorithms", Eidgenossische Technische Hochschule, Zurich, Switzerland, 2008.

[14] LJ. Stanković, M. Daković, S. Vujović, "Adaptive Variable Step Algorithm for Missing Samples Recovery in Sparse Signals," IET Signal Processing, vol. 8, no. 3, pp. 246 -256, 2014.

[15] I. Orovic, A. Draganic, S. Stankovic, "Sparse Time-Frequency Representation for Signals with Fast Varying Instantaneous Frequency," IET Radar, Sonar & Navigation, Volume: 9, Issue: 9, pp.: 1260 – 1267.

[16] Y. C. Eldar G. Kutyniok, "Compressed Sensing: Theory and Applications", Cambridge University Press, May 2012.

[17] S. Stankovic, I. Orovic, LJ. Stankovic, "An Automated Signal Reconstruction Method based on Analysis of Compressive Sensed Signals in Noisy Environment," Signal Processing, vol. 104, Nov 2014, pp. 43 - 50, 2014.

[18] Y. C. Eldar "Sampling Theory: Beyond Bandlimited Systems",Cambridge University Press, April 2015.

[19] I. Orovic, V. Papic, C. Ioana, X. Li, S. Stankovic, "Compressive Sensing in Signal Processing: Algorithms and Transform Domain Formulations," Mathematical Problems in Engineering, Review paper, 2016

[20] T. Blumensath, M. E. Davies, "Iterative Thresholding for Sparse Approximations", Journal of Fourier Analysis and Applications, vol. 14, no. 5-6, pp 629-654, December 2008

[21] LJ. Stankovic, S. Stankovic, M. Amin, "Missing Samples Analysis in Signals for Applications to L-estimation and Compressive Sensing," Signal Processing, vol. 94, Jan 2014, pp. 401-408, 2014.

[22] M. Brajovic, I. Orovic, M. Dakovic, S. Stankovic, "On the Parameterization of Hermite Transform with Application to the Compression of QRS Complexes," Signal Processing, Volume 131, February 2017, Pages 113-119.